\documentstyle[prl,aps,twocolumn]{revtex}

\begin{document}

\draft
\title{ Semiclassical quantization with short periodic orbits}
\author{ Eduardo G. Vergini and Gabriel Carlo} 

\address{ Departamento de F\'{\i}sica, Comisi\'on Nacional de
Energ\'{\i}a At\'omica. 
 Av. del Libertador 8250, 1429 Buenos Aires, Argentina.}

\date{\today}
\maketitle

\begin{abstract}
We apply a recently developed semiclassical theory of short peridic orbits
to the stadium billiard. We give explicit  expresions
for the resonances of periodic orbits and for the application
of the semiclassical Hamiltonian operator to them. Then, by using
the 3 shortest periodic orbits and 2 more living in the bouncing
ball region, we obtain the first 25 odd-odd eigenfunctions with
surprising accuracy. 

\end{abstract}

\centerline{PACS numbers: 05.45.Mt, 03.65.Sq, 45.05.+x}

%
%

\narrowtext
The study of semiclassical tecniques
in order to obtain quantum information of classically chaotic
Hamiltonian systems, has received much attention in the last 30 years
\cite{gut,bog,vor,ber,cita,sie,bog2};
most of the developed methods are
related to the Gutzwiller's trace formula \cite{gut}.
This formula is very attractive because gives the energy spectrum
of a bounded system in terms of periodic orbits ({\bf PO}s). 
However, the number
of {\bf PO}s required in the calculuation is enormous and increases
exponentially with the Heisenberg time 
$T_{H}\equiv 2 \pi \hbar \rho_{E}$ ($\rho_{E}$ is the mean energy density).

Recently, a new approach has been developed \cite{ver0} which uses a very little
number of short {\bf PO}s in the chaotic region. In order to verify 
the power of this new formalism, we applied it to the
Bunimovich stadium billiard with radius 
$R=1$ and straigth line $2$, an ergodic system \cite{buni}.
The starting point is the construction of resonances of a given 
unstable {\bf PO}.
They are functions (highly localized in energy) living in a 
neighborhood of the {\bf PO}. We will give the classical elements
in order to obtain explicit expresions for the resonances and for
the application of the semiclassical Hamiltonian operator to them.
Then, we select a set of resonances such that its mean density  
agrees with semiclassical prescriptions. Finally, we will evaluate
eigenfunctions and eigenvalues of the billiard by solving a generalized
eigenvalue problem.

Let $\gamma$ be a {\bf PO} of the
desymmetrized stadium billiard with turning points (a libration 
\cite{com}); see Fig. ~\ref{f1}.
Let $x$ be the coordinate along $\gamma$ with the origin $x=0$ at
one of the turning points; the other being at $x=L/2$, with $L$ the
length of $\gamma$. The transversal coordinate is $y$, with $y=0$ on
$\gamma$. 

Let $M(x)$ be a symplectic matrix describing the linearized 
transversal motion along $\gamma$; that is, a point with transversal
coordinates $(y,p_{y})$ at $x=0$ evolves according to the following rule:
$(y(x),p_{y}(x))=(y,p_{y})M(x)^{t}$. 
Then, $\tilde{M}(x)\equiv (-1)^{N(x)} M(x)$, with $N(x)$ the number of bounces
with the desymmetrized boundary while evolving from $0$
to $x$, is obtained with two types of matrices:
\[
M_{1}(l)=\left( \begin{array}{cc}
	    1   &   l  \\
	    0   &   1
		\end{array}  \right),\;\;{\rm and}\;\;\;
M_{2}(\theta)=\left( \begin{array}{cc}
	    1   &   0  \\
	    -2/ \cos(\theta)   &   1
		\end{array}  \right).
\]
$M_{1}(l)$ describes the evolution for a path of length $l$ without
bounces with the circle (the transversal momentum is measured in
units of the momentum along the trajectory), and $M_{2}(\theta)$
has into account a bounce with the circle ($\theta$ defines the
angle between the incoming trajectory and the radial direction).
$\sqrt{M_{2}(\theta)}$ is obtained from $M_{2}(\theta)$ by replacing
the $2$ by $1$. 

If the point $x=0$ (or $x=L/2$) is over the circle, we divide the
contribution $M_{2}(\theta)$ given by the bounce, between the incoming and outgoing path. For example (see Fig. ~\ref{f1}), 
$\tilde{M}(L/2)=M_{1}(\sqrt{5})\sqrt{M_{2}(\theta)}$ for orbit
(a), $M_{1}(1+\sqrt{2}) \sqrt{M_{2}(0)}$ for (b),
$M_{1}(\sqrt{3}) M_{2}(\pi/6) M_{1}(\sqrt{3}/2)$ for (c),
$\sqrt{M_{2}(0)} M_{1}(1+\sqrt{5})$ for (d),
$\sqrt{M_{2}(0)} M_{1}(1+\sqrt{10})$ for (e), and
$M_{1}(1/\sqrt{2}) M_{2}(\pi/4) M_{1}(1+1/\sqrt{2})$ for (f).

By using time reversal, it is easy to see that 
\[
\tilde{M}(L)=\left( \begin{array}{cc}
	    d  &   b  \\
	    c   &  a
		\end{array}  \right)
\left( \begin{array}{cc}
	    a   &   b  \\
	    c   &   d
		\end{array}  \right),
\]
where the explicit matrix on the right is $\tilde{M}(L/2)$ and the other completes
the orbit (from $x=L/2$ to $x=L\equiv 0$). Moreover, as the diagonal
elements of $\tilde{M}(L)$ are equal, the matrix can be written
as follows \cite{ver0}
\[ 
 \tilde{M}(L)=(-1)^{\nu} \left( \begin{array}{cc}
                \cosh (\lambda L)  &  \sinh (\lambda L)/\tan (\varphi) \\
 \sinh (\lambda L)\; \tan(\varphi) &   \cosh (\lambda L)
                                 \end{array}  \right).
\]        
$\lambda=(1/L) \ln(|A|+\sqrt{A^{2}-1})\;$ ($A\equiv ad+bc$) 
is the Lyapunov exponent in units of $[length^{-1}]$,
$\tan(\varphi)\;(\ne 0)$ in units of $[length^{-1}]$ defines the
slope of the unstable manifold in the plane $y-p_{y}$ 
(the slope of the stable manifold being $-\tan (\varphi)$), and
$\nu$ is the maximum number of conjugated points along $\gamma$. 
Finally, being $\xi_{u}$ and $\xi_{s}$
the unstable and stable directions respectively, the symplectic matrix $B$ transforming
coordinates from the new directions into the old ones ($y$ and $p_{y}$) is
\[ 
B=(\xi_{u}\;\xi_{s})= (1/ \sqrt{2}) \left( \begin{array}{cc}
			  	1/\alpha   &   -s/\alpha  \\
				s\; \alpha   &      \alpha 
				\end{array}  \right) ,
\]
with $\;\alpha\equiv\sqrt{|\tan (\varphi )|}=|ac/bd|^{1/4}$, 
and $s\equiv {\rm sign} (\varphi)={\rm sign}(acA)$.

Now, it is possible to construct a family of resonances associated to 
$\gamma$. Resonances within a family are identified by 
$n=0,1,\ldots$, the number of excitations along the trajectory,
and the wave number $k$ used in the construction depends on $\gamma$
and $n$ through the Bohr-Sommerfeld quantization rule:
\[
Lk-(N_{s}+s_{h}N_{h}+s_{v}N_{v}) \pi - \nu \pi/2=2 n\pi.
\]
$\nu$ is equal to the number of bounces with the circle, 
$N_{s}$ with the stadium boundary, and
$N_{h}$ ($N_{v}$) with the horizontal (vertical)
symmetry line. $s_{h}=0$ ($1$) for even (odd)
symmetry on the horizontal axis and equivalently with $s_{v}$ for
the vertical axis. $(\nu,N_{s},N_{h},N_{v})=(1,2,1,1)$
for (a), $(1,2,2,1)$ for (b), $(2,2,2,1)$ for (c), 
$(1,3,3,1)$ for (d), $(1,4,4,1)$ for (e), and $(2,2,1,1)$ for (f).

Resonances are constructed with straight lines by associating
a semiclassical expression to each one. The first line is defined
by the segment of $\gamma$ starting at $x_{1}=0$.
Let $x_{2}$ ($>x_{1}$) be the value of $x$ such that the path reaches the
stadium boundary while evolving along $\gamma$. The path going out
of $x_{2}$ defines the second line, and so on up to $x=L/2$.
It is necessary $1$ line for (a) and (b), and $2$ lines for (c), (d),
(e) and (f) (see Fig. ~\ref{f1}).

Defining local coordinates $(x^{(j)},y^{(j)})$ on each line such that
$x^{(j)}=x$ inside the desymmetrized billiard, the expresion for
line $j$ is [in the following expresions we are going to use $(x,y)$ 
understanding $(x^{(j)},y^{(j)})$]
\begin{equation}
\psi_{j}(x,y)=f_{j}(x,y)\; \sin[k y^{2} g_{j}(x)+kx-\Phi_{j}(x)],
\label{ec1}
\end{equation}
with 
$g_{j}(x)=[Q_{j}^{*}(x)P_{j}(x)+Q_{j}(x)P_{j}^{*}(x)]/(4\;|Q_{j}(x)|^{2})$,
$f_{j}(x,y)=2(k/ \pi)^{1/4} \exp[-ky^{2}/(2\; |Q_{j}(x)|^{2})]
/\sqrt{L|Q_{j}(x)|}$, and 
$\Phi_{j}(x)=\pi N_{D}(x_{j}^{+})+\varphi_{j}+[\phi_{j}(x)+\alpha_{j}(x)]/2$. 
$N_{D}(x_{j}^{+})$ is the
number of bounces up to $x_{j}$ (including the bounce at $x_{j}$)
satisfying Dirichlet boundary conditions; $N_{D}(x_{1}^{+})=0$. 
Moreover, 
\[
\left( \begin{array}{c} 
		Q_{j}(x) \\
		P_{j}(x)
	         \end{array} \right)
= M_{1}(x-x_{j}) \left( \begin{array}{cc}
	    a_{j}   &   b_{j}  \\
	    c_{j}   &   d_{j}
		\end{array}  \right)
\left( \begin{array}{c}
	       e^{-(x-x_{j})\lambda}   \\
	      i\; e^{(x-x_{j})\lambda} 
		\end{array}   \right),
\]
with
\[
\left( \begin{array}{cc}
	    a_{j}   &   b_{j}  \\
	    c_{j}   &   d_{j}
		\end{array}  \right)
=\tilde{M}(x_{j}^{+}) B \left( \begin{array}{cc}
	    e^{-x_{j}\lambda}   &   0  \\
	    0   &   e^{x_{j}\lambda}
		\end{array}  \right).
\]
$x_{j}^{+}$ means (for $j\!\geq \!2$) that $\tilde{M}$ is 
evaluated after the bounce with the boundary at $x_{j}$
[$\tilde{M}(x_{2}^{+})=M_{2}(\pi/6)M_{1}(\sqrt{3}/2)$ for
orbit (c), $M_{1}(\sqrt{5}/2)$ for (d), $M_{1}(2\sqrt{10}/3)$ for
(e), and $M_{2}(\pi/4)M_{1}(1+1/\sqrt{2})$ for (f)].
$\tilde{M}(x_{1}^{+})=\sqrt{M_{2}(\theta)}$ if $x_{1}$ lies on
the circle; otherwise $\tilde{M}(x_{1}^{+})=\openone$. 
$\phi_{j}(x)={\rm arg}[Q_{j}(x)]-{\rm arg}[Q_{j}(x_{j})]$ (arg takes
the argument of a complex number in the range $[0,2\pi)$). 
$\alpha_{j}(x)=2\pi\; {\rm sign}(x-x_{j})$ if  $n_{j}(x)\neq n_{j}(x_{j})$
and $(x-x_{j}) \phi_{j}(x)<0$; otherwise $\alpha_{j}(x)=0$, where
\[
n_{j}(x)=\left\{ \begin{array}{rl}
1 & {\rm if}\; x - x_j >{\rm max}(-a_{j}/c_{j},-b_{j}/d_{j}) \\
-1 & {\rm if}\; x - x_j <{\rm min}(-a_{j}/c_{j},-b_{j}/d_{j}) \\
0  & {\rm otherwise}       .    
					\end{array} \right.
\]
If $c_{j}\!=\!0$ or $d_{j}\!=\!0$, $x_{j}$ is replaced
by any other point on line $j$, inside the desymmetrized billiard.
$\phi_{j}(x)+\alpha_{j}(x)$ defines the angle swept
by $Q_{j}(x^{(j)})$ in a continuous way.    
Finally, 
$\varphi_{j}=\varphi_{j-1}+[\phi_{j-1}(x_{j})+\alpha_{j-1}(x_{j})]/2$
for $j \! \geq \!2$.
The value of $\varphi_{1}$ depends on the starting point and the symmetry,
and all the possibilities are considered in Fig. ~\ref{f1}. $\varphi_{1}$
is equal to $-s_{h}\pi/2$ for (a), $0$ for (b), $(s_{h}-1)\pi/2$
for (c), $(s_{h}+s_{v}-1)\pi/2$ for (d), $-s_{v}\pi/2$ for (e), and 
$(s_{v}-1)\pi/2$ for (f).

The transformation from local coordinates $(x^{(j)},y^{(j)})$ on line
$j$ to coordinates
$(X,Y)$ (horizontal and vertical directions in the plane 
respectively) is obtained through a simple 
transformation. If $(X_{j},Y_{j})$ are the coordinates of the point
$x_{j}$, and $\alpha_{j}$ the angle of line $j$ with the horizontal 
direction, $(x^{(j)}-x_j,y^{(j)})=G_{j}(X,Y)$ is given by
\[
G_{j}(X,Y)=(X-X_{j},Y-Y_{j}) 
\left( \begin{array}{cc}
	   \cos (\alpha_{j})   &  -\sin( \alpha_{j})  \\
	   \sin( \alpha_{j})   &  \cos (\alpha_{j})
		\end{array}  \right) .
\]
Finally, the family of resonances $\psi_{\gamma}(X,Y)$ is constructed 
with all the lines including symmetries (see Fig. ~\ref{f2}) 
\begin{equation}
\psi_{\gamma}=\sum_{j} \sum_{i=1}^{m_{h}} \sum_{l=1}^{m_{v}}
h_{i}\; v_{l}\; \psi_{j}[(x_j,0)+G_{j}(s_{l}X,s_{i}Y)].
\label{ec2}
\end{equation}
$s_{i}\equiv(-1)^{i+1}$ and $s_{l}\equiv(-1)^{l+1}$.
$h_{i}=[\delta_{i,1}+\delta_{i,2}(1-2s_{h})]$ and
$v_{l}=[\delta_{l,1}+\delta_{l,2}(1-2s_{v})]$. 
$m_{h}$ and $m_{v}$ depend on $j$ and are specified as follows:
$m_{h}=1$ ($2$)
if the line is (is not) symmetric with respect to the
horizontal axis, and equivalently with $m_{v}$ for the vertical
axis; however, $m_{h}=2$ and $m_{v}=1$ if the line goes through 
the origin.

We define $\hat{H}\equiv -\nabla^{2}$ and $E\equiv k^{2}$ 
(remember that $k$ depends on 
$\gamma$ and $n$). Then, the semiclassical approximation for
$(\hat{H}-E) \psi_{\gamma}(X,Y)$ is
obtained directly from Eq. (\ref{ec2}) having
into account the following semiclassical prescription \cite{ver0}
\begin{equation}
(\hat{H}\!-\!E)\psi_{j}(x,y)=\tilde{f_{j}}(x,y)\; 
\sin[k y^{2} g_{j}(x) +\Delta(x)],
\label{ec3}
\end{equation}
with 
$\tilde{f_{j}}(x,y)=\lambda k\;(2ky^{2}/|Q_{j}(x)|^{2}-1)f_{j}(x,y)$, 
and $\Delta(x)=kx-\Phi_{j}(x)+\pi/2-2 {\rm arg}[Q_{j}(x)]$. That is, the
action of $\hat{H}-E$ on $\psi_{j}$ excites the transversal direction
with two excitations. 
Using these expresions it is possible to obtain matrix elements
by direct integration on the domain (the quarter of billiard in
this case) \cite{surf}. 

In the semiclassical limit ($n \rightarrow \infty$), the 
following diagonal matrix elements are obtained explicitly
\cite{ver0} (using Dirac's notation):
$i)\;\langle \gamma | \gamma \rangle \rightarrow 1\;$, $ii)\;
\overline{E}\equiv  \langle \gamma | \hat{H} | \gamma \rangle/
 \langle \gamma | \gamma \rangle \rightarrow E \;$, and $iii)\;
 \sigma^{2}\equiv\langle \gamma|\hat{H}^{2}|\gamma 
\rangle/ \langle \gamma | \gamma \rangle\! -\! \overline{E}^{2}
 \rightarrow 2\lambda^{2} k^{2}$, with $\sigma$ the dispersion of
$\psi_{\gamma}$.
On the other hand, as the 
operator $\hat{H}$ is not exactly Hermitian, it is defined a symmetrized
interaction between {\bf PO}s as follows:
$\langle \delta |\hat{H} | \gamma \rangle \equiv 
 (\langle \delta |\hat{H} \gamma \rangle +
\langle \gamma |\hat{H} \delta \rangle ^{*})/2 $. 

Now we will select the set of resonances defining an
appropiate basis. From now on, we will restrict to odd symmetry on the
horizontal and vertical directions. For the selection of bouncing
ball functions we use Tanner's prescription \cite{tann}. That is,
for a given number $M$ ($M=1,2,\ldots$) of vertical excitations,
there are $N=[\sqrt{2M+1}+5/4]$ horizontal excitations in
the range $M < k/\pi<M+1$. This translates into a number $n$ of longitudinal 
excitations (along the selected orbit) that is specified by  
$n=(M-1)K+N-1$, where $K$ is the number of segments of the 
trajectory inside the desymmetrized billiard [for instance, 
$K=2$ for (b)]. Then, for the
construction of the function $(M,N)$ we select the trajectory
(among (b), (d) and (e)) which gives the resonance with 
smallest dispersion $\sigma$. 

The number of bouncing ball wavefunctions is only a fraction of the
mean number of states specified by the Weyl's law. Then, using 
the shortest periodic orbits (a), (b) and (c), we put as many
resonances as necessary to obtain the required number. As the period of
the three {\bf PO}s is comparable, we select first orbit (c)
because the associated resonances have smallest dispersion. 
Figure ~\ref{f3} clarifies the situation. The first column shows the
spectrum of bouncing ball resonances. Over each line
appears the label corresponding to Fig. ~\ref{f4}(a), $M$, $N$, 
the orbit used, and $n$. Second column shows
the spectrum of resonances constructed with orbit (c), and over
each line appears the label and $n$. The same is applicable to the 
third and fourth columns
with orbits (a) and (b) respectively. In this way, the density of 
resonances
agrees with the semiclassical mean density; however for $k> 15$
more orbits are required. Note that orbit (b) is used for the
construction of resonances living in the bouncing ball and
chaotic regions. This is because the bouncing ball region decreases
as $1/\sqrt{k}$.
Orbit (f) is not considered because the
associated resonances do not satisfy boundary conditions with 
sufficient accuracy for low energies.

Semiclassical eigenfunctions are constructed with this set of
wavefunctions. Of course, only a limited number of them are required
for a particular eigenfunction. State $21$ (see Fig. ~\ref{f4}(b)) needs
resonances from $18$ to $23$ (Fig. ~\ref{f4}(a)). 
The other states use less resonances. Suppose that at  
$k_{0}$ there is an eigenstate of the system;
it is constructed with all resonances satisfying 
$|k-k_{0}|\leq 0.8$ \cite{ver0}. Despite this is a 
semiclassical criterium,
it works in general at low energies too.
It says that the
number $N_{r}$ of resonances contributing to each eigenstate increases
as follows: $N_{r}\simeq 0.5\; k_{0}$.

Figure ~\ref{f4}(a) shows linear density plots of the 27 selected resonances
arranged by energy. Numbers below each plot are the label (left),
the squared root of the mean energy, 
and the dispersion $\sigma$ in units of the mean energy spacing.
Using this basis of functions, we evaluate the overlaps and 
the Hamiltonian matrix
elements using the semiclassical prescription given in (\ref{ec3}).
Then, by solving a generalized eigenvalue problem, the semiclassical
set of eigenfunctions shown in Fig. ~\ref{f4}(b) is obtained. Numbers below 
are the label, the semiclassical wave number and the overlap with
the exact solutions, which are displayed in Fig. \ref{f4}(c).
The mean standard deviation of the semiclassical eigenvalues is a fraction ($0.06$)
of the mean level spacing in accordance with the theory. Recent results 
obtained for the hyperbola billiard \cite{Keat} 
show a standard deviation of 0.047 (in units of the mean 
level spacing) for the first 24 even eigenvalues. Those results were 
calculated using trace-formula-type techniques, involving 38 131 periodic 
orbits (193 695 pseudo-orbits), in contrast with the only 5 used in the 
present work. On the other hand, 
overlaps of the semiclassical eigenfunctions with the exact ones
are surprisingly good (see Fig ~\ref{f4}(b)). Moreover, to our knowledge, 
this is the first semiclassical evaluation of a set of eigenfunctions in a
chaotic system using periodic orbits. 

In conclusion we have applied successfully the theory of short 
{\bf PO}s to the stadium billiard. This shows that the classical
information contained in short {\bf PO}s is sufficient for obtaining
the stationary states of a bounded chaotic Hamiltonian system.  
 
This work was partially supported by PICT97 03-00050-01015, SECyT-ECOS and 
CONICET PIP 0420/98. E.V. acknowledges the hospitality of the LPTMS, Orsay, 
where part of this work was done.

\begin{figure}
\caption{ Set of periodic orbits of the desymmetrized stadium billiard
used (with the exception of orbit (f)) for the construction of resonances.
(1) and (2) labels the corresponding straight line. }
\label{f1}
\end{figure}

\begin{figure}
\caption{ Set of lines including symmetries used for the construction
of resonances associated to orbit (d). The different sets of coordinates
used are indicated. }
\label{f2}
\end{figure}

\begin{figure}
\caption{ Spectrum of resonances used for the evaluation of semiclassical
eigenfunctions. The meaning of each column is as follows: the first number 
over the lines is the label. In the first column, showing the spectrum of 
bouncing ball resonances, $M$, $N$ and $n$ are indicated. Letters correspond 
to  the orbit from which they were constructed. In the other three columns 
only the label and the number $n$ is over the lines; the 
corresponding orbit, the same for the entire column, is at the bottom of 
the figure.}
\label{f3}
\end{figure}

\begin{figure}
\caption{ Linear density plots of: (a) the selected resonances. The numbers 
below each plot are (from left to right) the label, the squared root of the 
mean energy, and the dispersion $\sigma$ in units of the mean energy spacing. 
(b) the semiclassical eigenfunctions. Here, after the label, the semiclassical 
wave number and the overlap with the corresponding exact eigenfunction are 
displayed. (c) exact eigenfunctions with their wavenumbers.}
\label{f4}
\end{figure}


\begin{references}
\bibitem{com} Rotations are required in the stadium billiard for
the evaluation of highly excited states.
\bibitem{surf} It is possible to obtain explicit expressions for the matrix 
elements working on a surface of section.
\end{references}
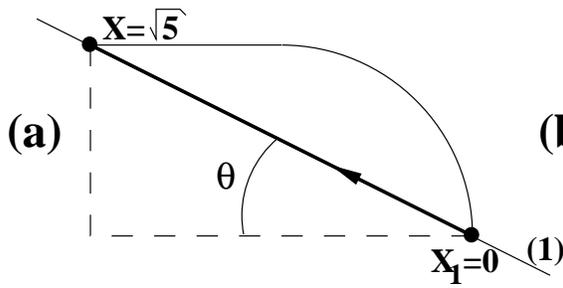
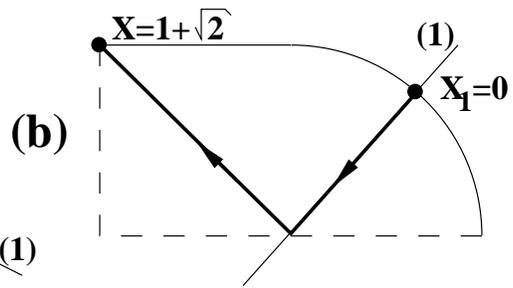
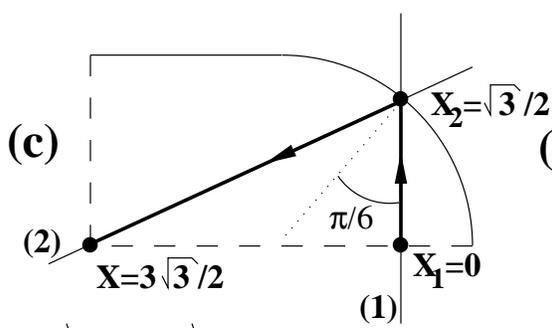
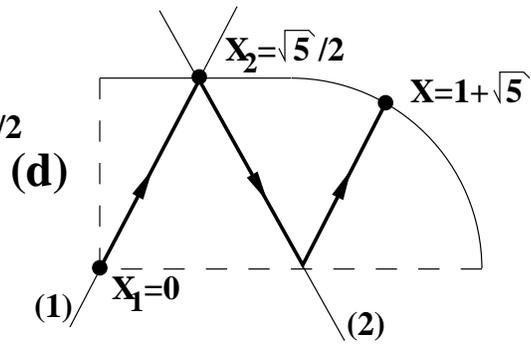
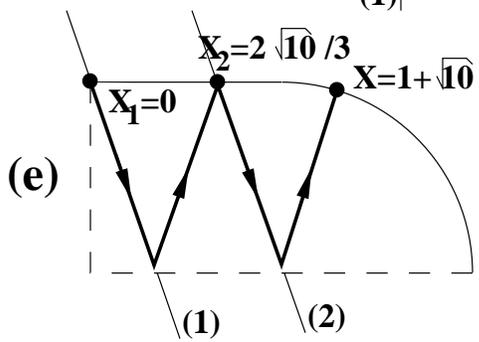
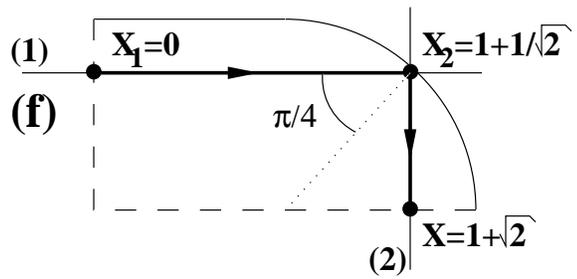
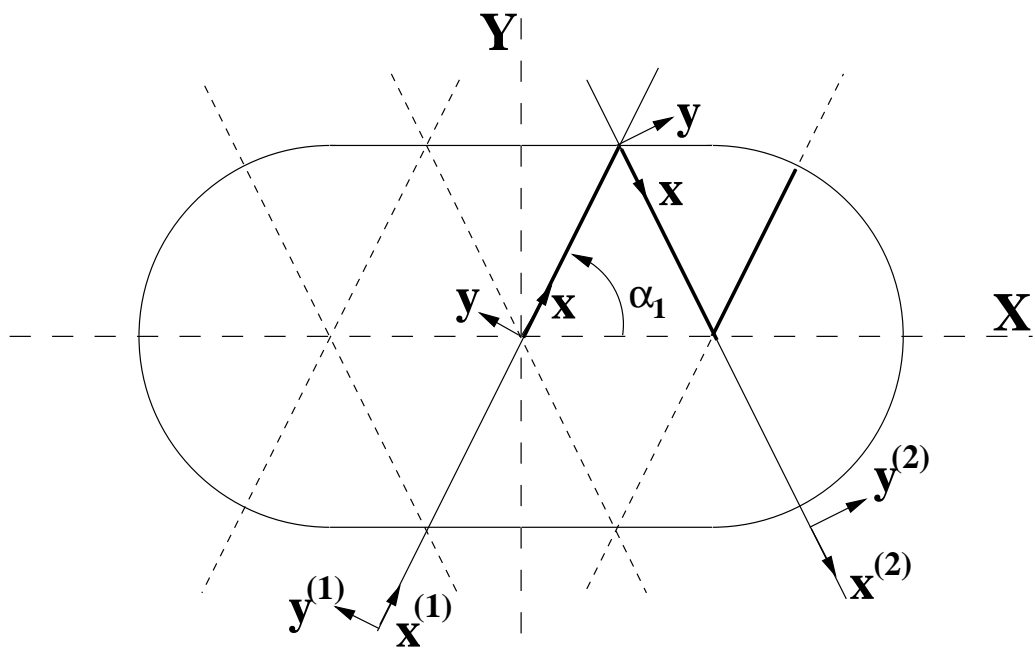
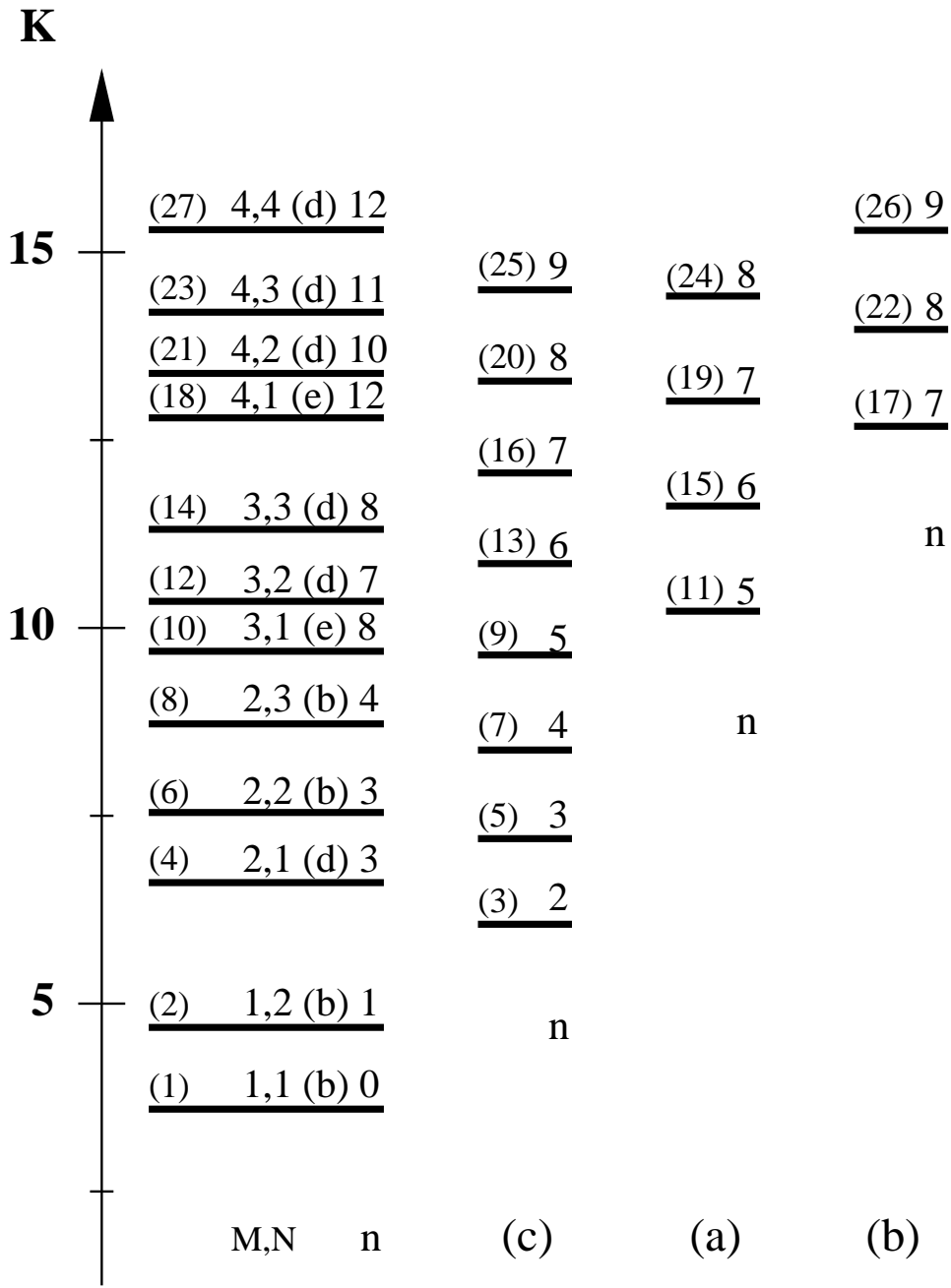
\end{document}